\documentclass[aps,prl,preprint,onecolumn,longbibliography]{revtex4-1}

\usepackage{fancyhdr}
\usepackage{graphicx}
\usepackage{amsmath,amsthm,amssymb}
\usepackage{float}
\usepackage[hidelinks=true]{hyperref}
\usepackage{subcaption}
\usepackage{color,soul} 

\newcommand{\bd}[1]{\textbf{#1}}
\renewcommand{\vec}[1]{\boldsymbol{#1}}

\begin{document}
\title{Inertial Effects on the Stress Generation of Active Fluids}
\author{S.C. Takatori\footnote{Present address: Miller Institute for Basic Research in Science, University of California, Berkeley, 2536 Channing Way, D-104, Berkeley, California 94720, USA; Department of Bioengineering, University of California, Berkeley, California 94720, USA} and J.F. Brady}	
\affiliation{Division of Chemistry and Chemical Engineering, California Institute of Technology, Pasadena, California 91125, USA}

\date{\today}

\begin{abstract}
Suspensions of self-propelled bodies generate a unique mechanical stress owing to their motility that impacts their large-scale collective behavior.  For microswimmers suspended in a fluid with negligible particle inertia, we have shown that the virial `swim stress' is a useful quantity to understand the rheology and nonequilibrium behaviors of active soft matter systems.  For larger self-propelled organisms like fish, it is unclear how particle inertia impacts their stress generation and collective movement.  Here, we analyze the effects of finite particle inertia on the mechanical pressure (or stress) generated by a suspension of self-propelled bodies.  We find that swimmers of all scales generate a unique `swim stress' and `Reynolds stress' that impacts their collective motion.  We discover that particle inertia plays a similar role as confinement in overdamped active Brownian systems, where the reduced run length of the swimmers decreases the swim stress and affects the phase behavior.  Although the swim and Reynolds stresses vary individually with the magnitude of particle inertia, the sum of the two contributions is independent of particle inertia.  This points to an important concept when computing stresses in computer simulations of nonequilibrium systems---the Reynolds and the virial stresses must both be calculated to obtain the overall stress generated by a system.
\end{abstract}

\maketitle
\sloppy

\section{Introduction}
Active matter systems constitute an intriguing class of materials whose constituents have the ability to self-propel, generate internal stress, and drive the system far from equilibrium.  Because classical concepts of thermodynamics do not apply to nonequilibrium active matter, recent work has focused on invoking the mechanical pressure (or stress) as a framework to understand the complex dynamic behaviors of active systems \citep{Takatori14,Takatori16a,Ginot15,Yang14,Mallory14,Solon15}.  In particular, active swimmers exert a unique ``swim pressure'' as a result of their self propulsion \citep{Takatori14,Yang14}.  A physical interpretation of the swim pressure is the pressure (or stress) exerted by the swimmers as they interact with the surrounding boundaries that confine them, similar to molecular or colloidal solutes that collide into the container walls to exert an osmotic pressure.  The swim pressure is a purely entropic, destabilizing quantity \citep{Takatori15a} that can explain the self-assembly and phase separation of a suspension of self-propelled colloids into dilute and dense phases that resemble an equilibrium gas-liquid coexistence \citep{Theurkauff12,Palacci13,Redner13,Bialke13,Stenhammar13,Cates10}.  

Existing studies of the pressure of active systems \citep{Takatori14,Takatori14b,Takatori15a,Yang14,Mallory14,Ginot15,Solon15} have focused on overdamped systems where swimmer inertia is neglected (i.e, the particle Reynolds number is small).  However, the swim pressure has no explicit dependence on the body size and may exist at all scales including larger swimmers (e.g., fish, birds) where particle inertia is not negligible \citep{Takatori14}.  The importance of particle inertia is characterized by the Stokes number, $St_R \equiv (M/\zeta)/\tau_R$, where $M$ is the particle mass, $\zeta$ is the hydrodynamic drag factor, and $\tau_R$ is the reorientation time of the active swimmer.  Here we analyze the role of nonzero $St_R$ on the mechanical stress exerted by a system of self-propelled bodies and provide a natural extension of existing pressure concepts to swimmers with finite inertia.  We maintain negligible fluid inertia so that the fluid motion satisfies the steady Stokes equation.  

We consider a suspension of self-propelled spheres of radii $a$ that translate with an intrinsic swim velocity $\vec{U}_0$ and tumble with a time $\tau_R$ in a continuous Newtonian fluid.  The random tumbling results in a diffusive process for time $t \gg \tau_R$ with $D^{swim} = U_0^2 \tau_R / 6$ in 3D.  An isolated active particle generates a swim pressure $\Pi^{swim} = n \zeta D^{swim} = n \zeta U_0^2 \tau_R/6$, where $n$ is the number density of active particles.  We do not include the effects of hydrodynamic interactions, and there is no macroscopic polar order of the swimmers or any large-scale collective motion (e.g., bioconvection). 
	
From previous work \citep{Takatori16b,Takatori16a,Yan15b,Ezhilan15,Tailleur09}, we know that geometric confinement of overdamped active particles plays a significant role in their dynamics and behavior.  Confinement from potential traps, physical boundaries, and collective clustering can reduce the average run length and swim pressure of the particles.  We have shown experimentally \citep{Takatori16b} that active Brownian particles trapped inside a harmonic well modifies the swim pressure to $\Pi^{swim} =  (n \zeta U_0^2 \tau_R/2) \left(1 + \alpha \right)^{-1}$ in 2D, where $\alpha \equiv U_0 \tau_R / R_c$ is a ratio of the run length, $U_0 \tau_R$, to the size of the trap, $R_c$.  For weak confinement, $\alpha \ll 1$, we obtain the `ideal-gas' swim pressure of an isolated swimmer.  For strong confinement, $\alpha \gg 1$, the swim pressure decreases as $\Pi^{swim}/(n \zeta U_0^2 \tau_R / 2) \sim 1/\alpha$.  Confinement reduces the average distance the swimmers travel between reorientation events, which results in a decreased swim pressure.

In this work, we find that particle inertia plays a similar role as confinement by reducing the correlation between the position and self-propulsive swim force of the swimmers.  In addition to the swim pressure, active swimmers exert the usual kinetic or Reynolds pressure contribution associated with their average translational kinetic energy.  For swimmers with finite particle inertia we find that the sum of the swim and Reynolds pressures is the relevant quantity measured by confinement experiments and computer simulations.  We also study systems at finite swimmer concentrations and extend our existing mechanical pressure theory to active matter of any size or mass.
	
An important implication of this work pertains to the computation of mechanical stresses of colloidal suspensions at the appropriate level of analysis.  Consider the Brownian osmotic pressure of molecular fluids and Brownian colloidal systems, $\Pi^B = n k_BT$, where $n$ is the number density of particles and $k_BT$ is the thermal energy.  At the Langevin-level analysis where mass (or inertia) is explicitly included, the Reynolds stress is the source of the Brownian osmotic pressure, $- \langle \rho \vec{U}' \vec{U}' \rangle$, where $\rho$ is the density and $\vec{U}'$ is the velocity fluctuation.  The virial stress or the moment of the Brownian force, $- n \langle \vec{x} \vec{F}^B \rangle$, is identically equal to zero---the position and Brownian force are uncorrelated at the Langevin level.  However, at the overdamped Fokker-Planck or Smoluchowski level where inertia is not explicitly included, the position and Brownian force are correlated, and the virial stress $- n \langle \vec{x} \vec{F}^B \rangle$ is the source of the Brownian osmotic pressure\cite{Brady93}, whereas the Reynolds stress is zero.  The important point is that the sum of the Reynolds and virial stresses gives the correct Brownian osmotic pressure at both levels of analysis, as it must be since the osmotic pressure of colloidal Brownian suspensions is $\Pi^B = n k_BT$ whether or not inertia is explicitly included in the analysis.

We report in this work that an identical concept applies for active swimmers.  The virial stress arising from the correlation between the particle position and its `internal' force, $- n \langle \vec{x} \vec{F} \rangle$, is a term that is separate and in addition to the Reynolds stress associated with their average translational kinetic energy.  Interestingly, the mechanical stress generated by active swimmers has a nonzero contribution from both the Reynolds and virial stresses because the internal force associated with self-propulsion has an autocorrelation that is not instantaneous in time and instead decays over a finite timescale modulated by the reorientation time of the active swimmer, $\tau_R$:  $\langle \vec{F}^{swim}(t) \vec{F}^{swim}(0) \rangle \sim e^{-2 t/\tau_R}$.  A distinguishing feature of active swimmers compared with passive Brownian particles is that their direction of self-propulsion can relax over large timescales, and that the `internal' swim force autocorrelation cannot in general be described by a delta-function in time.

\section{Swim stress}	
All self-propelled bodies exert a swim pressure, a unique pressure associated with the confinement of the active body inside a bounded domain.  The swim pressure is the trace of the swim stress, which is defined as the symmetric first moment of the self-propulsive force, $\vec{\sigma}^{swim} = - n \langle \vec{x} \vec{F}^{swim} \rangle^{sym}$, where $n$ is the number density of swimmers, $\vec{x}$ is the position, $\vec{F}^{swim}$ is the swimmer's self-propulsive swim force, and $\langle \cdot \rangle ^{sym}$ denotes the symmetric part of the tensor.  As we noted previously\cite{Takatori17a}, the swim stress is properly defined as the symmetric force moment since the force arises from the fluid which can only generate symmetric stresses.  For the active Brownian particle model, the swim force can be written as $\vec{F}^{swim} = \zeta U_0 \vec{q}$ where $\vec{q}$ is a unit vector specifying the swimmer's direction of self-propulsion.  For a dilute suspension of active particles with negligible particle inertia the ``ideal-gas" swim stress is given by $\vec{\sigma}^{swim} = - n \zeta U_0^2 \tau_R \vec{I} / 6 = -n k_s T_s \vec{I}$, where we define $k_s T_s \equiv \zeta U_0^2 \tau_R / 6$ as the swimmer's ``energy scale'' (force ($\zeta U_0$) $\times$ distance ($U_0\tau_R$)).  The swim pressure (or stress) is entropic in origin and is the principle destabilizing term that facilitates a phase transition in active systems \citep{Takatori15a}.

In the absence of any external forces, the motion of an active Brownian particle is governed by the Langevin equations:
\begin{align}
M \frac{\partial \vec{U}}{\partial t} &= - \zeta (\vec{U} - \vec{U}_0) + \sqrt{2 \zeta^2 D_0} \vec{\Lambda}_T, \label{eq_Chap8:1} \\
I \frac{\partial \vec{\Omega}}{\partial t} &= - \zeta_R \vec{\Omega} + \sqrt{\frac{2 \zeta_R^2}{\tau_R}} \vec{\Lambda}_R, \label{eq_Chap8:2} 
\end{align}
where $M$ and $I$ are the particle mass and moment-of-inertia, $\vec{U}$ and $\vec{\Omega}$ are the translational and angular velocities, $\zeta_R$ is the hydrodynamic drag factor coupling angular velocity to torque,  $\sqrt{2 \zeta^2 D_0} \vec{\Lambda}_T$ and $\sqrt{2 \zeta_R^2/\tau_R} \vec{\Lambda}_R$ are the Brownian translational force and rotational torque, respectively, $\vec{\Lambda}_T$ and $\vec{\Lambda}_R$ are unit random normal deviates, $\tau_R \sim 1/D_R$ is the reorientation timescale set by rotational Brownian motion, and $D_0$ is the Stokes-Einstein-Sutherland translational diffusivity.  The translational diffusivity and the reorientation dynamics are modeled with the usual white noise statistics, $\langle \Lambda_i(t) \rangle = 0$ and $\langle \Lambda_i(t) \Lambda_j(0) \rangle = \delta(t) \delta_{ij}$.  The swimmer orientation $\vec{q}(t)$ is related to the angular velocity by the kinematic relation $\vec{\Omega} \times \vec{q} = d \vec{q} / dt$.  The translational and angular velocities may be combined into a single vector, $\mathcal{U} = (\vec{U},\vec{\Omega})^T$, and similarly for the force and torque, $\mathcal{F} = (\vec{F},\vec{L})^T$, to obtain a general solution to the system of ordinary differential equations \citep{Hinch75}.  Although a general solution is available for any particle mass and moment-of-inertia, inclusion of nonzero moment-of-inertia leads to calculations that are analytically involved.  For convenience and to make analytical progress, here we summarize the case of zero moment-of-inertia ($I = 0$) and focus on finite mass ($St_R \equiv (M/\zeta)/\tau_R \ne 0$) to elucidate the effects of inertia on the dynamics of active matter.  

We can solve Eqs \ref{eq_Chap8:1} and \ref{eq_Chap8:2} for the swimmer configuration ($\vec{x}(t), \vec{q}(t)$), and calculate the swim stress.  As shown in the Appendix, the swim stress for arbitrary particle inertia is
\begin{equation} \label{eq_Chap8:3}
\vec{\sigma}^{swim} = - n k_s T_s \left( \frac{1}{1 + 2 St_R} \right) \vec{I},
\end{equation}
where we have taken times $t > \tau_M$ and $t > \tau_R$, $\tau_M \equiv M/\zeta$ is the swimmer momentum relation time, energy scale $k_s T_s \equiv \zeta U_0^2 \tau_R / 6$, and $St_R \equiv \tau_M / \tau_R = (M/\zeta)/\tau_R$ is the Stokes number.  For $St_R = 0$ we recover the ``ideal-gas" swim pressure for an overdamped system: $\Pi^{swim} = - \text{tr} \vec{\sigma}^{swim}/3 = n \zeta U_0^2 \tau_R / 6 = n k_s T_s$ \citep{Takatori14}.  This is precisely the mechanical force per unit area that a dilute system of confined active micro-swimmers exert on its surrounding container \citep{Takatori14,Yang14,Mallory14}.  

Notice in the other limit as $St_R \rightarrow \infty$, $\vec{\sigma}^{swim}$ vanishes.  Physically, the magnitude of the swim stress decreases because inertia may translate the swimmer in a trajectory that is different from the direction of the swim force, reducing the correlation $\langle \vec{x} \vec{F}^{swim} \rangle$ between the moment arm $\vec{x}$ and the orientation-dependent swim force $\vec{F}^{swim} = \zeta U_0 \vec{q}$.  Our earlier work \citep{Takatori16b} showed that active particles confined by an acoustic trap exert a swim pressure that is reduced by a factor of $(1 + \alpha)^{-1}$, where $\alpha \equiv U_0 \tau_R / R_c$ is the degree of confinement of the swimmer run length relative to the size of the trap $R_c$. Equation \ref{eq_Chap8:3} has a reduction in the swim pressure by a similar factor, $(1 + 2 St_R)^{-1}$, suggesting that particle inertia may play a similar role as confinement by reducing the correlation between the position and self-propulsive direction of the swimmers.  

Particle inertia may be interpreted as imposing a confinement effect on the swimmers because their effective run length between reorientation events decreases. The average run length of the swimmers with inertia reduces to $\sim (U_0 \tau_R - \Delta x')$, where $\Delta x' = U_0 \Delta t$ is the distance over which inertia translates the swimmer along a trajectory that is independent of the direction of its swim force, and the time over which this occurs scales with the inertial relaxation time, $\Delta t \sim M/\zeta$.  Substituting these terms into the virial expression for the swim pressure, we obtain $\Pi^{swim} \sim n \langle \vec{x} \cdot \vec{F}^{swim} \rangle \sim n \zeta U_0^2 \tau_R (1 - (M/\zeta)/\tau_R)$.  Using the definition of the Stokes number $St_R \equiv (M/\zeta)/\tau_R$ and for small $St_R$, we can rewrite the swim pressure as $\Pi^{swim} \sim n \zeta U_0^2 \tau_R (1 + St_R)^{-1}$.  Aside from the factor of $2$ in the denominator (which arises from spatial dimensionality), this scaling argument agrees with Eq \ref{eq_Chap8:3} and shows that particle inertia plays a confining role in the swim pressure, analogous to the physical confinement of swimmers in a potential well\citep{Takatori16b}.  

\begin{figure}[t!]
	\centering
	\includegraphics[width=0.6\textwidth]{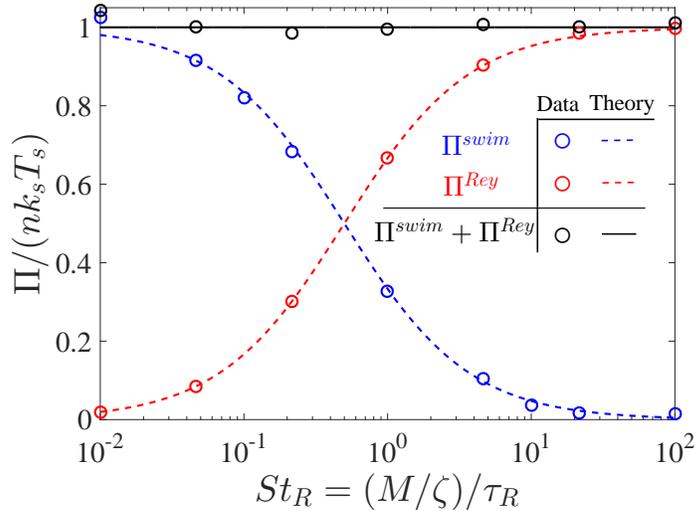}
	\caption{Swim and Reynolds pressures of a dilute system of swimmers with finite inertia, where $\Pi = - \text{tr} \vec{\sigma} / 3$.  The blue ($\Pi^{swim}$) and red ($\Pi^{Rey}$) curves and symbols are the analytical theory of Eqs \ref{eq_Chap8:3} and \ref{eq_Chap8:4} and simulation data, respectively.  The solid black line is the sum of the swim and Reynolds stresses.  The Brownian osmotic pressure $\Pi^B = n k_B T$ has been subtracted from $\Pi^{Rey}$.}  \label{Fig_Chap8:1}
\end{figure}

\section{Reynolds stress}
For systems with finite particle inertia, an additional stress contribution arises owing to particle acceleration: the Reynolds stress.  This term is seen in Bernoulli's equation and is associated with the average translational kinetic energy of a particle, $\vec{\sigma}^{Rey} = -n M \langle \vec{U}' \vec{U}' \rangle$.  In atomic or molecular systems this is often referred to as the kinetic stress.  This contribution was not included in previous studies since overdamped active systems have no particle mass, $M=0$ (i.e., $St_R \equiv (M/\zeta)/\tau_R = 0$).  As shown in the Appendix, we can use the solution to Eqs \ref{eq_Chap8:1} and \ref{eq_Chap8:2} to obtain the Reynolds stress for arbitrary $St_R$, given by
\begin{equation} \label{eq_Chap8:4}
\vec{\sigma}^{Rey} = - n k_B T \vec{I} - n k_s T_s \left( \frac{1}{1 + 1/(2 St_R)} \right) \vec{I},
\end{equation}
which is a sum of the Brownian osmotic stress $\vec{\sigma}^B = - n k_B T \vec{I}$ and a self-propulsive contribution that depends on $St_R$.  For an overdamped system where $St_R = 0$, the self-propulsive contribution to the Reynolds stress vanishes, justifying the neglect of this term in previous studies of overdamped systems.  Mallory et al \citep{Mallory14} analytically calculated the expression of the Reynolds stress, Eq \ref{eq_Chap8:4}, but not the swim stress, Eq \ref{eq_Chap8:3}; the mechanical pressure that is measured from the walls of an enclosing container is the sum of the Reynolds and swim pressures, however.

Notice that the Brownian osmotic stress, $\vec{\sigma}^B = -n k_B T\vec{I}$, arises solely from the Reynolds stress and not from taking the virial moment of the Brownian force, $\langle \vec{x} \vec{F}^B \rangle$.  As stated earlier, this is precisely because of the delta-function statistics imposed for the Brownian force in the Langevin-level analysis where mass is explicitly included: $\langle \vec{F}^B(t) \vec{F}^B(0) \rangle = 2 k_B T \zeta \delta(t) \vec{I}$.  In contrast, the active stress has a nonzero contribution from both the Reynolds and swim stresses because the swim force autocorrelation is a decaying exponential modulated by the reorientation timescale $\tau_R$:  $\langle \vec{F}^{swim}(t) \vec{F}^{swim}(0) \rangle \sim e^{-2 t/\tau_R}$.  If one were to model the Brownian force autocorrelation as one that relaxes over a finite solvent relaxation timescale, $\langle \vec{F}^B(t) \vec{F}^B(0) \rangle \sim e^{-2t/\tau_S}$, then we would obtain nonzero contributions from both the Reynolds and virial stresses, with their sum equal to $\vec{\sigma}^B = -n k_B T \vec{I}$ for all $\tau_S$.

For a dilute suspension, the stress exerted by an active swimmer is the sum of the swim and Reynolds stresses.  Adding Eqs \ref{eq_Chap8:3} and \ref{eq_Chap8:4}, we find 
\begin{equation} \label{eq_Chap8:5}
\vec{\sigma}^{swim} + \vec{\sigma}^{Rey} = - n (k_B T + k_s T_s) \vec{I}. 
\end{equation}
Remarkably, the Stokes number $St_R$ disappears.  The magnitude of the swim pressure that decreases with increasing $St_R$ cancels exactly the increase in the magnitude of the Reynolds stress.  This verifies that swimmers of all scales exert the pressure, $nk_s T_s$, regardless of their mass and inertia.  We conducted simulations where the dynamics of active Brownian particles were evolved following Eqs \ref{eq_Chap8:1} and \ref{eq_Chap8:2} using the velocity verlet algorithm \citep{Allen89}, and the results are shown in Fig \ref{Fig_Chap8:1}.  Results from the simulations agree with our theoretical predictions in Eqs \ref{eq_Chap8:3} - \ref{eq_Chap8:5}.  In an experiment or simulation, the average mechanical pressure exerted on a confining boundary gives the sum of the swim and Reynolds pressures, and not their separate values.  

We have shown previously that random motion gives rise to a micromechanical stress via the relationship $\vec{\sigma} = - n \zeta \vec{D}$, where $\vec{D}$ is the effective translational diffusivity\cite{Takatori14}.  For overdamped active systems, the self-propulsive contribution to the diffusivity is described solely through the swim stress,  $\vec{\sigma}^{swim} = - n \zeta \vec{D}^{swim} = -n \zeta U_0^2 \tau_R \vec{I} / 6$.  With finite particle inertia, the particle diffusivity is unaltered and the stress-diffusivity relationship still applies but the stress is now a sum of two independent contributions, as shown in Eq \ref{eq_Chap8:5}.  

In the presence of a nonzero moment-of-inertia, there is another dimensionless Stokes number, $St_I \equiv (I/\zeta_R)/\tau_R$, which is a ratio of the inertial reorientation timescale, $\tau_I = I/\zeta_R$, and the swimmers' intrinsic reorientation timescale, $\tau_R$.  Similar to the translational Stokes number $St_R$ that does not appear in Eq \ref{eq_Chap8:5}, the moment-of-inertia is not expected to appear explicitly in the total stress generated by an active swimmer; at long times the random walk diffusive motion is unaffected by the particle inertia, whether that be translational or rotational.

\begin{figure} 
	\includegraphics[width=0.495\textwidth]{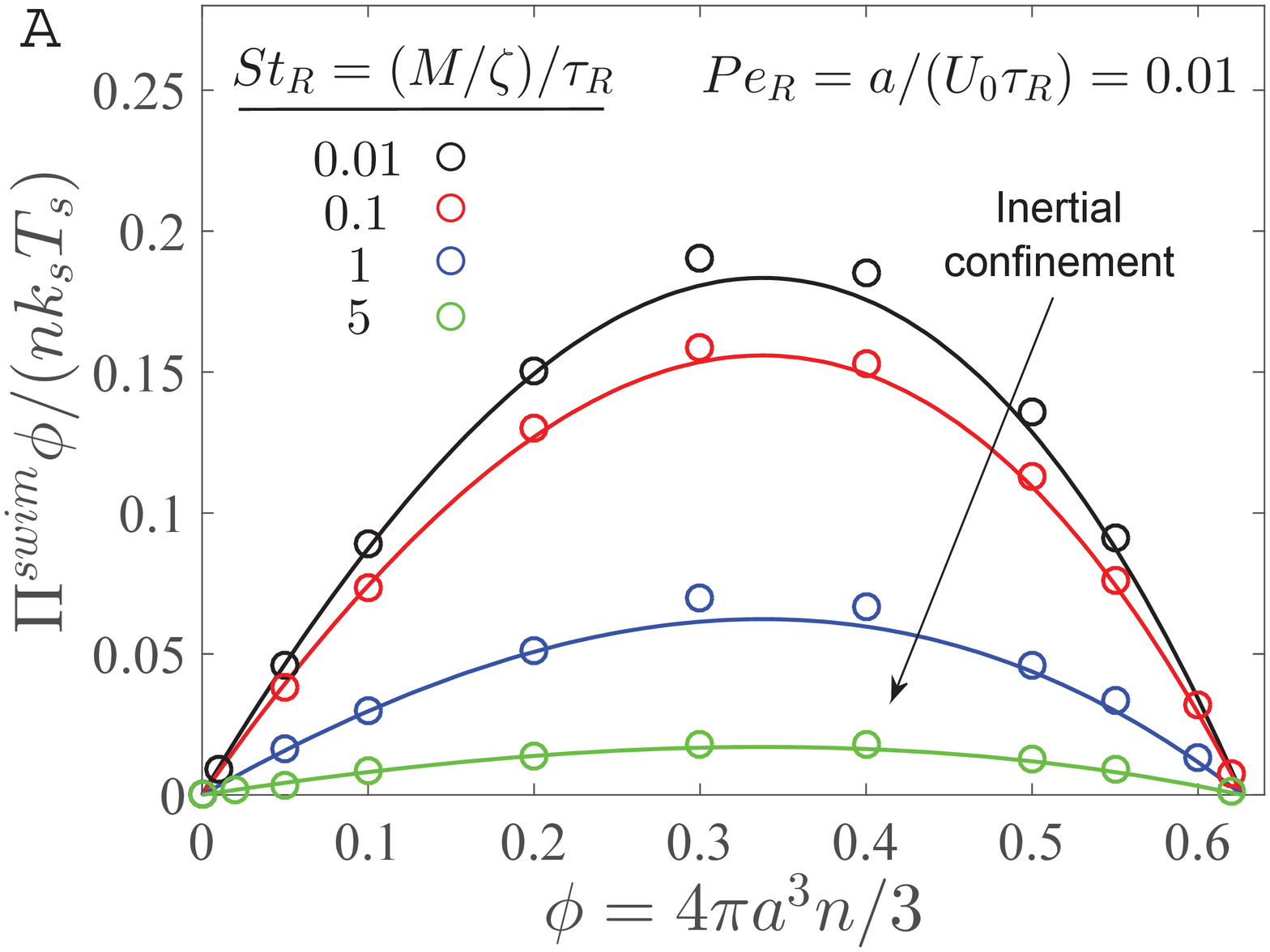}	
	\includegraphics[width=0.495\textwidth]{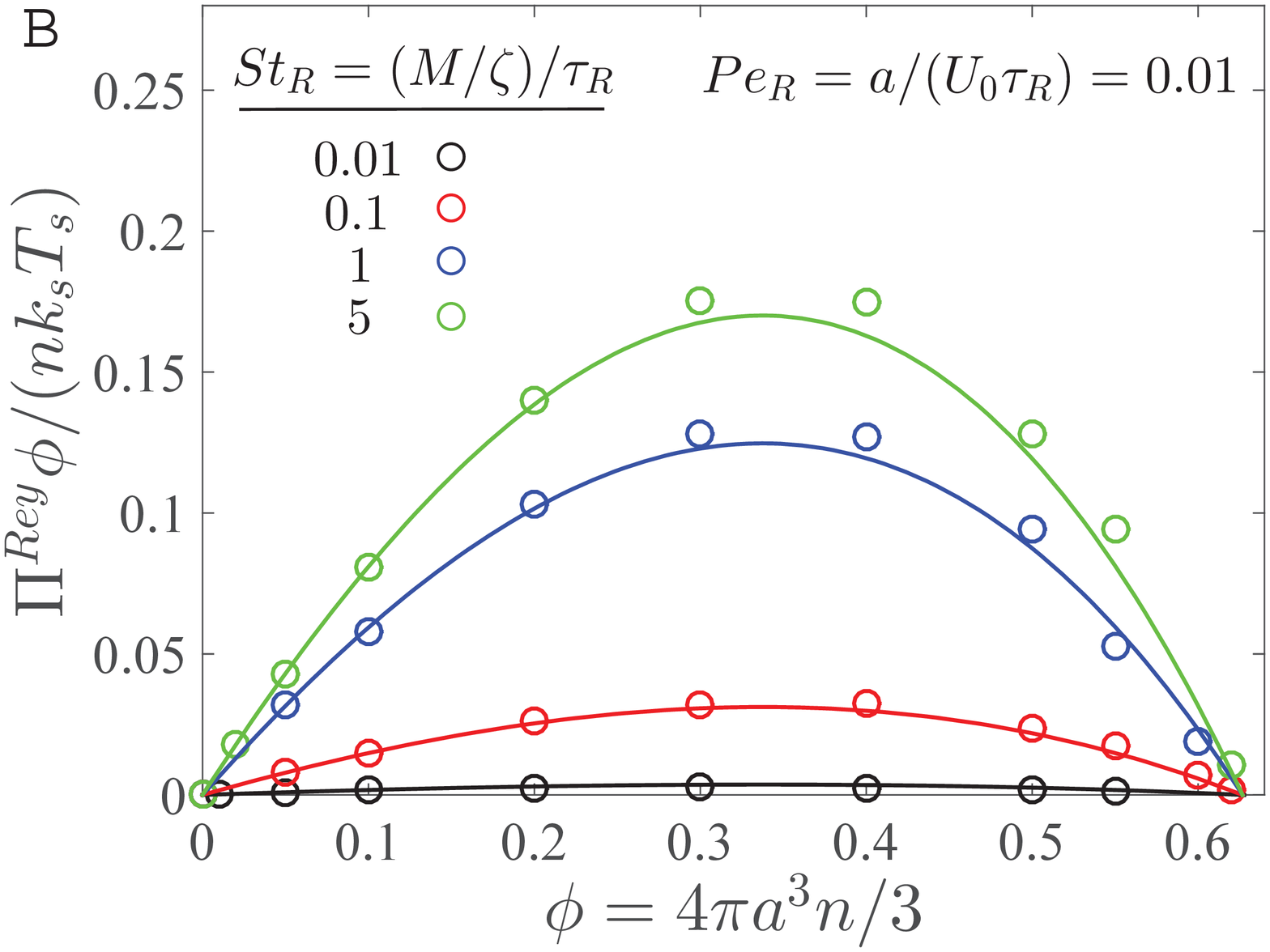}
	\caption{(\bd{A}) Swim pressure, $\Pi^{swim}$, and (\bd{B}) Reynolds pressure, $\Pi^{Rey}$, as a function of volume fraction of particles, $\phi$, for different values of $St_R \equiv (M/\zeta)/\tau_R$ and a fixed reorientation P\'eclet number, $Pe_R \equiv a/(U_0\tau_R) = 0.01$.  The symbols and solid curves are the simulation data and analytical theory, respectively.  The Brownian osmotic pressure $\Pi^B = n k_B T$ has been subtracted from the Reynolds pressure.} \label{Fig_Chap8:2}
\end{figure}

\section{Finite concentrations}
The results thus far are for a dilute suspension of active swimmers.  At finite concentrations, experiments and computer simulations have observed unique phase behavior and self-assembly in active matter \citep{Theurkauff12,Palacci13,Redner13,Bialke13,Stenhammar13,Cates10}.  Recently a new mechanical pressure theory was developed to provide a phase diagram and a natural extension of the chemical potential and other thermodynamic quantities to nonequilibrium active matter \citep{Takatori15a}.

At finite concentrations of swimmers, dimensional analysis shows that the nondimensional swim and Reynolds stresses depend in general on $(St_R, \phi, Pe_R, k_s T_s/(k_B T))$, where $\phi = 4 \pi a^3 n / 3$ is the volume fraction of active swimmers and $Pe_R \equiv a/(U_0 \tau_R)$ is the reorientation P\'eclet number---the ratio of the swimmer size $a$ to its run length $U_0\tau_R$.  The ratio $k_s T_s/(k_B T)$ quantifies the magnitude of the swimmers' activity ($k_s T_s \sim \zeta U_0^2 \tau_R$) relative to the thermal energy $k_BT$; this ratio can be a large quantity for typical micro-swimmers.

From previous work on overdamped active systems with negligible particle inertia \citep{Takatori14}, we know that $Pe_R$ is a key parameter controlling the phase behavior of active systems.  For large $Pe_R$ the swimmers reorient rapidly and take small swim steps, behaving as Brownian walkers; the swimmers thus do not clump together to form clusters and the system remains homogeneous.  For small $Pe_R$ the swimmers obstruct each others' paths when they collide for a time $\tau_R$ until they reorient.  This decreases the run length of the swimmers between reorientation events and causes the system to self-assemble into dense and dilute phases resembling an equilibrium liquid-gas coexistence.

As reported previously for $St_R = 0$ \citep{Takatori14,Takatori15a}, for small $Pe_R$ the swim pressure decreases with increasing swimmer concentration.  To verify how finite particle inertia affects the swim pressure at larger concentrations, we conducted simulations by evolving the motion of active particles following Eqs \ref{eq_Chap8:1} and \ref{eq_Chap8:2}, with an additional hard-sphere interparticle force $\vec{F}^P$ that prevents particle overlaps using a potential-free algorithm \citep{Heyes93}.  Care was taken to ensure that the simulation time step was small enough to preclude unwanted numerical errors associated with resolution of particle collisions.  We varied the simulation time step from $dt/\tau_R = 10^{-5} - 10^{-3}$ and found a negligible difference in our results.  As shown in Fig \ref{Fig_Chap8:2}A, for finite $St_R$ the data from our simulations are well described by the expression $\Pi^{swim} = n k_s T_s (1 - \phi - \phi^2) / (1 + 2 St_R)$, which is simply a product of a volume fraction dependence and a Stokes number dependence of Eq \ref{eq_Chap8:3}.  The volume fraction dependence $(1 - \phi - \phi^2)$ was used previously to model the phase behavior of active matter \citep{Takatori15a}.

The clustering of swimmers reduces their translational velocity autocorrelation, $\langle \vec{U}' \vec{U}' \rangle$, and hence decreases the Reynolds pressure.  As shown in Fig \ref{Fig_Chap8:2}B, our simulations show that the Reynolds pressure decreases with concentration, increases with $St_R$, and is well described by the expression $\Pi^{Rey} = n k_s T_s (1 - \phi - \phi^2) / (1 + 1/(2 St_R)) + n k_B T$. 

\begin{figure} 
	\centering
	\includegraphics[width=0.6\textwidth]{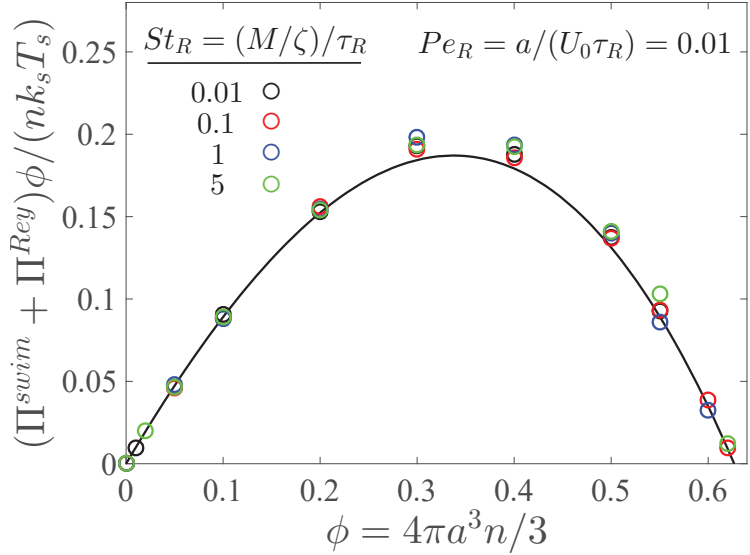}
	\caption{Sum of swim and Reynolds pressures, $\Pi^{swim} + \Pi^{Rey} = - \text{tr} (\vec{\sigma}^{swim} + \vec{\sigma}^{Rey}) / 3$, as a function of volume fraction of particles $\phi$ for different values of $St_R \equiv (M/\zeta)/\tau_R$ and a fixed reorientation P\'eclet number, $Pe_R \equiv a/(U_0\tau_R) = 0.01$.  The symbols and solid curve are the simulation data and analytical theory of Eq \ref{eq_Chap8:6}, respectively.  The Brownian osmotic pressure $\Pi^B = n k_B T$ has been subtracted from the total pressure.}  \label{Fig_Chap8:3}
\end{figure}

The sum of the swim and Reynolds stresses is given by
\begin{equation} \label{eq_Chap8:6}
\Pi^{swim} + \Pi^{Rey} = n k_B T + n k_s T_s (1 - \phi - \phi^2),
\end{equation}
which again has no dependence on $St_R$ (nor $Pe_R$ for $Pe_R < 1$ considered here), even at finite $\phi$.  Equation \ref{eq_Chap8:6} is corroborated by our simulations as shown in Fig \ref{Fig_Chap8:3}.  This result implies that the existing mechanical pressure theory \citep{Takatori15a} developed for overdamped systems can be used directly for swimmers with finite Stokes numbers, as long as we include the Reynolds stress contribution into the active pressure.  Inclusion of the Reynolds stress is critical, as confinement experiments and computer simulations measure the total active pressure, including both the swim and Reynolds contributions.  

\begin{figure} 
	\centering
	\includegraphics[width=0.6\textwidth]{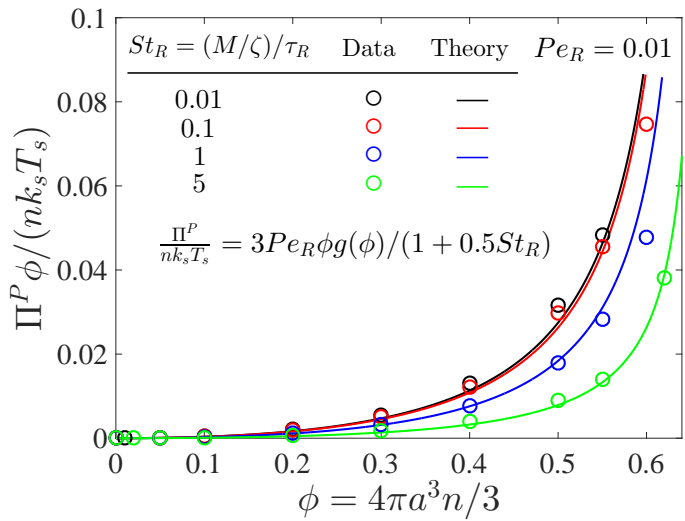}
	\caption{Interparticle collisional pressure, $\Pi^P$, as a function of volume fraction of particles, $\phi$, for different values of $St_R \equiv (M/\zeta)/\tau_R$ and a fixed reorientation P\'eclet number, $Pe_R \equiv a/(U_0\tau_R) = 0.01$.  The symbols and solid curves are the simulation data and analytical expression, respectively.}  \label{Fig_Chap8:4}
\end{figure}

In addition to the swim and Reynolds stresses, interparticle interactions between the swimmers at finite concentrations give rise to an interparticle stress, $\vec{\sigma}^P(St_R, \phi, Pe_R, k_s T_s/(k_B T))$.  For repulsive interactions, the interparticle (or collisional) pressure, $\Pi^P =  - \text{tr} \vec{\sigma}^P / 3$, increases monotonically with concentration and helps to stabilize the system.  As shown in Fig \ref{Fig_Chap8:4}, we find that the expression $\Pi^P / (n k_s T_s) = 3 Pe_R \phi g(\phi) / (1+0.5 St_R)$ agrees with the simulation data for a fixed value of $Pe_R = 0.01$, where $g(\phi) = \left( 1 - \phi/\phi_0 \right)^{-1}$ is the pair distribution function at particle contact, and $\phi_0 = 0.65$ is a parameter obtained from the interparticle pressure of hard-sphere molecular fluids \citep{Takatori15a}.  We can add $\Pi^P$ to Eq \ref{eq_Chap8:6} to construct phase diagrams of a system of inertial swimmers, which are qualitatively similar to those presented in  \citep{Takatori15a} for small values of $St_R$.  Adding particle inertia shifts the stablization to larger $\phi$.

As shown by Batchelor \citep{Batchelor70}, there may be an additional contribution to the particle stress arising from local fluctuations in acceleration, $\vec{f}'$, and is given by $ -(1/V) \sum \int_{V_p} \rho \vec{f}' \vec{r} dV$, where $V$ is the volume of the suspension (fluid plus particles), $V_p$ is the volume of an individual particle, $\rho$ is the uniform density of the particle, $\vec{r}$ is a position vector (or the moment arm) from the particle center, and the summation is over the number of particles in the volume $V$.  For a dilute system of rigid particles, this term arises only from solid body rotation of the particle and takes the form $\int_{V_p} \vec{f}' \vec{r} dV = (4 \pi a^5 /15) (\vec{\Omega}_P \vec{\Omega}_P - \vec{\Omega}_P \cdot \vec{\Omega}_P \vec{I})$, where $\vec{\Omega}_P$ is the average angular velocity of the rigid particle of size $a$.  Here the active swimmers have no average angular velocity, so there is no stress arising from local fluctuations in acceleration for dilute active systems of rigid particles.

In addition to using a potential-free algorithm to model hard-sphere particles, we have also tested a short-ranged, repulsive Weeks-Chandler-Andersen potential with an upper cut-off at particle separation distances of $r = 2^{1/6}(2a)$.  Using this softer potential, the swim pressure does not exhibit a concentration dependence of $(1 - \phi - \phi^2)$ because the effective radius of the particles decreases as the system becomes denser, meaning that colliding particles exhibit increasingly large overlaps.  Increasing particle inertia (i.e., larger $St_R$) also changed the effective particle size.  As previously stated \citep{Stenhammar14}, one must use care when soft potentials are used to model hard-sphere particle collisions because the effective particle size may depend on system parameters (like $Pe_R$ and $St_R$).

\section{Conclusion}
Here we presented a mechanical pressure theory for active Brownian particles with finite inertia.  We neglected hydrodynamic interactions between the swimmers, which may contribute additional terms (like the ``hydrodynamic stresslet" \citep{Saintillan08a}) to the active pressure.  The ratio of the magnitudes of the hydrodynamic stress to the swim stress is $\sigma^H / \sigma^{swim} \sim (n \zeta U_0 a)/(n \zeta U_0^2 \tau_R) = a/(U_0 \tau_R) \equiv Pe_R$.  The hydrodynamic stress contribution becomes negligible when phase-separation occurs at low $Pe_R$.

We assumed that the surrounding fluid obeys the steady Stokes equations, which may not be true for larger swimmers that propel themselves using fluid inertia.  However, the concepts of the swim and Reynolds stress apply for swimmers with a nonlinear hydrodynamic drag factor, $\zeta_{inertia}(|\vec{U}|)$, where $|\vec{U}|$ is the magnitude of the swimmer velocity.  For example, a self-propelled body may experience a fluid drag that is quadratic in the velocity $\vec{F}^{drag} \sim \zeta_{inertia}(|\vec{U}|) \vec{U} \sim (\rho_s a^2 |\vec{U}|) \vec{U}$, where $\rho_s$ is the fluid density and $a$ is the characteristic size of the body.  The nondimensional Langevin equations would become $d\vec{U}/dt = - A |\vec{U}| (\vec{U} - \vec{q})$, where $\vec{q}$ is the orientation vector of the swimmer and $A = \rho_s a^2 U_0 \tau_R / M \sim (1/St_R)(\zeta_{inertia}/(\eta a))$ is the relevant quantity that must be varied.

\section{acknowledgments}
The authors thank Luis Nieves-Rosado for his contributions to this work as part of the Undergraduate Research Program at the California Institute of Technology.  S.C.T. acknowledges support by the Gates Millennium Scholars fellowship and the National Science Foundation Graduate Research Fellowship under Grant No. DGE-1144469. This work was also supported by NSF Grant CBET 1437570.

\section{Appendix}
Integrating Eq \ref{eq_Chap8:1} twice in time, we obtain the position of the swimmer,
\begin{multline} \label{eq_Chap8:A1}
\vec{x}(t) = \vec{x}(0) + \vec{U}(0) \tau_M \left( 1 - e^{-t/\tau_M} \right) + \\
\int_0^t \left( U_0 \vec{q}(t') + \sqrt{2 D_0} \vec{\Lambda}_T(t') \right) (1 - e^{-(t-t')/\tau_M}) \ dt',
\end{multline}
where $\vec{x}(0)$ and $\vec{U}(0)$ are the arbitrary initial position and velocity, respectively, $\tau_M \equiv M/\zeta$ is the momentum relaxation time, $U_0$ is the intrinsic swimmer velocity, $\vec{q}$ is the unit orientation vector of the swimmer, and $\vec{\Lambda}_T$ is a unit random deviate.  Integrating Eq \ref{eq_Chap8:2} and using the kinematic relation, $\vec{\Omega} \times \vec{q} = d \vec{q} / dt$, we obtain
\begin{equation} \label{eq_Chap8:A2}
\frac{d \vec{q}}{dt} = \vec{\Omega}(0)\times \vec{q}(t) e^{-t/\tau_I} + \frac{1}{\tau_I} \sqrt{\frac{2}{\tau_R}} \int_0^t \vec{\Lambda}_R(t') \times \vec{q}(t') e^{-(t-t')/\tau_I} \ dt',
\end{equation}
where $\vec{\Omega}(0)$ is the initial angular velocity, $\tau_I = I/\zeta_R$ is the angular momentum relaxation time, and $\vec{\Lambda}_R$ is a unit random deviate.  Equation \ref{eq_Chap8:A2} is of the form $d q_i/dt = A_{ik}(t) q_k$, where $A_{ik}(t)$ is a coefficient matrix.  The general solution of Eq \ref{eq_Chap8:A2} is $q_i(t) = q_k(0) e^{\int_0^t A_{ik}(t') \ dt'}$, where $q_k(0)$ is an arbitrary initial orientation of the swimmer.

We are interested in the orientation autocorrelation
\begin{equation} \label{eq_Chap8:A3}
\langle q_i(t) q_n(t') \rangle = \frac{1}{3} \delta_{nk} \langle e^{\int_{t'}^t A_{ik}(t'') dt''} \rangle,
\end{equation}
where $A_{ik}(t) = \epsilon_{ijk} \left( \Omega_j(0) e^{-t/\tau_I} + \sqrt{2/(\tau_R \tau_I^2)} \int_0^t \Lambda_j(t') e^{-(t-t')/\tau_I} dt' \right)$ is the coefficient matrix, and $\vec{\epsilon}$ is the unit alternating tensor.  In the limit of small angular momentum relaxation time, $\tau_I \rightarrow 0$, we obtain $A_{ik}(t) = \sqrt{2/\tau_R} \epsilon_{ijk} \Lambda_j(t)$, and 
\begin{equation} \label{eq_Chap8:A4}
\langle q_i(t) q_n(t') \rangle = \frac{1}{3} \delta_{in} e^{-2(t-t')/\tau_R}.
\end{equation}
Notice that as $\tau_R \rightarrow 0$ the autocorrelation becomes a delta function and the swimmer reorients rapidly and behaves as a Brownian walker.

Using Eqs \ref{eq_Chap8:A1}, \ref{eq_Chap8:A4}, and the swim force $\vec{F}^{swim} \equiv \zeta U_0 \vec{q}$, the swim stress is given by
\begin{multline} \label{eq_Chap8:A5}
\vec{\sigma}^{swim} = - n \langle \vec{x} \vec{F}^{swim} \rangle^{sym} = \frac{n}{3} \zeta U_0^2 \vec{I} \left[ \frac{\tau_R}{2} \left( 1 - e^{-2t/\tau_R} \right) - \right. \\ \left.
\frac{1}{2/\tau_R + 1/\tau_M} \left( 1 - e^{-(2/\tau_R + 1/\tau_M)t} \right) \right],
\end{multline}
where we have used that $\langle \vec{x}(0) \vec{F}^{swim}(t) \rangle = \langle \vec{U}(0) \vec{F}^{swim}(t) \rangle = \langle \vec{F}^{swim}(t') \vec{\Lambda}_T(t) \rangle = \vec{0}$.  Taking times $t > \tau_M$ and $t > \tau_R$, we obtain Eq \ref{eq_Chap8:3} of the main text.

Following a similar procedure, the Reynolds stress is given by
\begin{multline} \label{eq_Chap8:A6}
\vec{\sigma}^{Rey} = - n M \langle \vec{U}' \vec{U}' \rangle = -n M \langle \vec{U}(0) \vec{U}(0) \rangle e^{-2t/\tau_M} - \\
\frac{nM}{3} \left(\frac{U_0}{\tau_M}\right)^2 \vec{I} \left\{ \frac{1}{2/\tau_R + 1/\tau_M} \left[ \frac{\tau_M}{2} \left( 1 - e^{-2t/\tau_M} \right) - \right. \right. \\ \left. \left. \frac{1}{- 2/\tau_R + 1/\tau_M} \left( e^{-(2/\tau_R + 1/\tau_M)t} - e^{-2t/\tau_M} \right) \right] + \right. \\ \left. 
\frac{1}{-2/\tau_R + 1/\tau_M} \left[ - \frac{\tau_M}{2} \left( 1 - e^{-2t/\tau_M} \right) + \frac{1}{2/\tau_R + 1/\tau_M} \left(1 - e^{-(2/\tau_R + 1/\tau_M)t} \right) \right]  \right\} - \\
n k_B T \left(1 - e^{-2t/\tau_M} \right) \vec{I}.
\end{multline}
Taking times $t > \tau_M$ and $t > \tau_R$, we obtain the Reynolds stress as given in Eq \ref{eq_Chap8:4} of the main text.

\bibliographystyle{unsrt}
\bibliography{Diffusion}

\end{document}